\documentclass[aps,pra,showpacs,superscriptaddress,twocolumn]{revtex4}
\usepackage{color}
\usepackage{amsmath}
\usepackage{graphicx}
\begin{document}
\title{Sturmian-Floquet approach to high-order harmonic generation}
\author{J\'{o}zsef Kasza}
\affiliation{ELI-ALPS, ELI-HU Non-profit Ltd., Dugonics t\'{e}r 13, H-6720 Szeged, Hungary}
\email{Jozsef.Kasza@eli-alps.hu}
\author{P\'{e}ter Dombi}
\affiliation{ELI-ALPS, ELI-HU Non-profit Ltd., Dugonics t\'{e}r 13, H-6720 Szeged, Hungary}
\affiliation{MTA "Lend\"{u}let" Ultrafast Nanooptics Group, Wigner Research Center for Physics, Konkoly-Thege Mikl\'{o}s \'{u}t 29-33, H-1121 Budapest, Hungary}
\author{P\'{e}ter F\"{o}ldi}
\affiliation{ELI-ALPS, ELI-HU Non-profit Ltd., Dugonics t\'{e}r 13, H-6720 Szeged, Hungary}
\affiliation{Department of Theoretical Physics, University of Szeged, Tisza Lajos k\"{o}r\'{u}t 84, H-6720 Szeged, Hungary}

\begin{abstract}
 We show that the Floquet approach with a
Sturmian basis means an efficient description of high-order harmonic generation with monochromatic excitation. This method, although involves numerical calculations, is close to analytic approaches with the corresponding deeper insight into the dynamics. As a first application, we investigate the role of atomic coherence during the process of HHG: as it is shown, different coherent superpositions of initial atomic states produce observably different HHG spectra. For linearly polarized excitation, we demonstrate that the question whether the constituents of the initial superpositions are dipole coupled or not, strongly influences the dynamics. By investigating time-dependent HHG signals, we also show that the preparation of the initial atomic state can be used for the control of the high-harmonic radiation.
\end{abstract}
\maketitle
\section{Introduction}

High-order harmonic generation (HHG) is a key concept in the recent
development of optics, partially because of its fundamental aspects
that can deepen our knowledge on nonlinear light-matter interaction \cite{KI09},
but also because of its importance in producing short, attosecond
range bursts of electromagnetic radiation \cite{FarkasT92,Agostini2004}. This process
was first observed for the case of gas samples \cite{McPherson87,F88}, but since
the beginning of the 90's, a wide variety of physical systems were
shown or predicted to produce high harmonics, including plasma surfaces
\cite{TG09,Tsakiris2006,Kahaly2013,Baeva2006}, bulk solids \cite{Ghimire10} and nanostructures \cite{Ciappina2014}.

In the current paper, we focus on the interaction of short, intense
laser fields and hydrogen-like atoms. In this context, the light-induced
dynamics of the electrons naturally involve the continuum part of
the spectrum, since the strong electromagnetic field of the laser
can induce an ionization process, i.e., it can transfer the electrons
to the continuum. In more detail, the most often used, so-called "three-step"
model \cite{0953-4075-24-15-004,C93b,Krause1992} consists of the emission
of a single active electron, its motion \char`"outside the atom",
in the laser field, and recombination with the parent ion. This model,
besides providing and instructive theoretical interpretation, can
describe the main features of the gas HHG spectra.

The full quantum mechanical
description of the process of HHG from gas samples means the solution
of the corresponding time-dependent Schr\"{o}dinger equation (TDSE) \cite{levenstein}.
There are various numerical methods for this purpose, having an important
point in common: the truncation of the underlying Hilbert-space. For
discretization in real space, it means using a sufficiently large,
but necessarily finite computational box, in order to avoid unphysical
reflections at the boundaries. As a different approximation, we can
try to use a finite set of the hydrogenic eigenstates to describe
the problem, or, as an intermediate way, can discretize in the radial
direction, and use spherical harmonic expansion \cite{Krause1992} for the remaining
two dimensions.

For sinusoidally oscillating external fields, Floquet's
method \cite{F883,PS89} offers a convenient way to transfer
the TDSE to a static eigenvalue problem. (Also for solid states systems, see e.g. \cite{FK97,SPF15}.) Numerically necessary truncation
in this case means ignoring harmonics above a given order. Recalling
the typical properties of gas HHG spectra (plateau and then a cutoff for increasing
frequencies), this approximation can safely be performed, numerical
errors can be kept under a predefined limit. The power and transparency
of this method is most obvious when not too many quantum mechanical
states have to be used to describe the time-dependent problem.

However, since a realistic description of the HHG process should involve the
continuum as well, usual hydrogen-like eigenstates are not adequate.
Instead, we use appropriate Sturmian functions \cite{Goscinski,Goscinski2002}, which cover both the
bound and the continuum part of the spectrum, and provide an appropriate
basis. As we show, the combination of the Floquet approach with a
Sturmian basis \cite{joachain2012atoms} means a method which is
close to analytic approaches with the corresponding deeper insight
into the dynamics. As a first application, we calculate HHG spectra for
various laser parameters. We pay special attention to the initial
atomic state, consider various superpositions of low-lying energy
eigenstates \cite{ayadi}. As we show, atomic coherence has strong influence on
the HHG signals, and this effect can be used for the optimization
of attosecond pulse generation.

In the following, first we introduce the theoretical framework, Floquet's method and the Sturmian basis in Sec.~\ref{theosec}. We describe our results in Sec.~\ref{resultsec} and draw the conclusions in Sec.~\ref{summarysec}.

\section{Floquet approach and Sturmian functions}

\label{theosec} The time dependent Schr\"{o}dinger equation that
we consider reads:
\begin{equation}
i\hbar\frac{d}{dt}|\psi\rangle(t)=[H_{0}+V(t)]|\psi\rangle(t),\label{tdse}
\end{equation}
where $H_{0}$ corresponds to the free atomic system (kinetic term plus
the Coulomb potential). The interaction with a linearly polarized,
monochromatic external laser field $\mathbf{E}$ is given by
\begin{equation}
V(t)=\mathbf{DE_{0}}\cos(\omega t),
\end{equation}
where $\mathbf{D}$ is the dipole moment operator and $\mathbf{E_{0}}$
is assumed to be parallel with the $x$ axis.

Since the Hamiltonian appearing in Eq.~(\ref{tdse}) is periodic
in time ($T=2\pi/\omega$), the infinite set of Floquet states
\begin{equation}
|\phi_{k}\rangle(t)=e^{i\epsilon_{k}t}|u_{k}\rangle(t)
\label{Fldef}
\end{equation}
with $|u_{k}\rangle(t+T)=|u_{k}\rangle(t)$ exists, and plays an important
role in the theoretical description of the problem. Specifically,
since the states $|\phi_{k}\rangle$ with $k=0,1,2,\ldots$ form a
proper basis for all values of $t,$ when the expansion
\begin{equation}
|\psi\rangle(0)=\sum_{k}c_{k}|\phi_{k}\rangle(0)=\sum_{k}c_{k}|u_{k}\rangle(0)\label{initsate}
\end{equation}
of an initial state is known, its time evolution can be calculated
as simply as
\begin{equation}
|\psi\rangle(t)=\sum_{k}c_{k}|\phi_{k}\rangle(t)=\sum_{k}c_{k}e^{i\epsilon_{k}t}|u_{k}\rangle(t).\label{evolution}
\end{equation}
That is, having determined the Floquet quasienergies $\epsilon_{k}$
and the corresponding time-periodic states $|u_{k}\rangle(t),$ we
essentially know the dynamics of any initial state.

Practically, $\epsilon_{k}$ and $|u_{k}\rangle(t)$ can be found
using a set of orthogonal functions in real space (like, e.g., the
usual hydrogenic eigenstates, $|n,l,m\rangle$) and considering Fourier
functions $e^{ik\omega t}$ as a time domain basis for the periodic
functions (with $k$ being an integer). Using spherical coordinates,
we can define
\begin{align}
& |n,l,m;k\rangle(\mathbf{r},t)=|n,l,m\rangle(\mathbf{r})e^{ik\omega t}=2^{l+1}e^{-\frac{r}{a_{0}n}}\left(\frac{r}{a_{0}n}\right)^{l}\nonumber \\
 & \times\sqrt{\frac{(-l+n-1)!}{a_{0}^{3}n^{4}(l+n)!}}\mathrm{L}_{-l+n-1}^{2l+1}\left(\frac{2r}{a_{0}n}\right)\mathrm{Y}_{l}^{m}(\theta,\phi)e^{ik\omega t},
 \label{nlmk}
\end{align}
where $L$ and $Y$ denote associated Laguerre polynomials and the
spherical harmonics, respectively. For the sake of simplicity, in
the following the arguments $(\mathbf{r},t)$ will not appear explicitly
in the notation, unless it is necessary. The states above are orthogonal
in terms of the inner product
\begin{equation}
\langle\varphi_{1}|\varphi_{2}\rangle=\int_{0}^{T}dt\int_{-\infty}^{\infty}d^{3}\mathbf{r}\varphi_{1}^{*}(\mathbf{r},t)\varphi_{2}(\mathbf{r},t),\label{innerprod}
\end{equation}
that is, $\langle n,l,m;k|n',l',m';k'\rangle=\delta_{nn'}\delta_{ll'}\delta_{mm'}\delta_{kk'}.$
Using Eq.~(\ref{Fldef}), the Floquet states
can be searched in the following form:
\begin{equation}
\left\vert \phi_{k}\right\rangle (t)=\exp(i\epsilon_{k}t)\sum\limits _{nlmk}c_{nlm}^{k}|n,l,m;k\rangle(t).\label{expand psi}
\end{equation}
Substituting this form back to the time dependent Schr\"{o}dinger equation
(\ref{tdse}), and multiplying the result from the left by $\langle n',l',m';k'|,$
we obtain;
\begin{equation}
\epsilon_{k'}c_{n'l'm'}^{k'}=\sum\limits _{nlm}c_{nlm}^{k}\langle n',l',m';k|H|n,l,m;k\rangle+k\omega\delta_{nn'}\delta_{ll'}\delta_{mm'}\delta_{kk'}.\label{Feigen}
\end{equation}
By denoting all the four discrete indices $(n,l,m,k)$ by a single
integer variable $j$, and introducing $\tilde{H}_{jj'}=\langle n',l',m';k|H|n,l,m;k\rangle+k\omega\delta_{nn'}\delta_{ll'}\delta_{mm'}\delta_{kk'},$
Eq.~(\ref{Feigen}) is seen to have the form of an eigenvalue equation:
\begin{equation}
\epsilon_{j}c_{j}=\sum\limits _{j'}c_{j'}\langle j|\tilde{H}|j'\rangle.\label{Feigen2}
\end{equation}
(Note that here, if $j$ corresponds to the indices $(n,l,m,k)$,
$c_{j}$ means $c^{nlm}_{k}$ and $\epsilon_{j}$ denotes $\epsilon_{k}.$)
The matrix elements appearing above can be determined using the coordinate
representation of the Hamiltonian $H(t)$ as well as the states $|n,l,m;k\rangle$
as given by Eq.~(\ref{nlmk}). Clearly, we have
\begin{equation}
\langle n',l',m';k'|H_{0}|n,l,m;k\rangle=\frac{\mathcal{E}_{0}}{n^{2}}\delta_{nn'}\delta_{ll'}\delta_{mm'}\delta_{kk'},\label{H0elements}
\end{equation}
where $\mathcal{E}_{0}=-1Ry\approx-13.6eV$ for the hydrogen atom.
The interaction term $V(t)$ has explicit time dependence, and consequently
will not be diagonal in the (time-related) last index:
\begin{align}
& \mathbf{DE_{0}}\cos(\omega t)|n,l,m;k\rangle=\frac{E_{0}}{2}D(e^{i\omega t}+e^{-i\omega t})|n,l,m;k\rangle\label{Velements}\\
 & =\frac{E_{0}}{2}D\left[|n,l,m;k+1\rangle+|n,l,m;k-1\rangle\right],\nonumber
\end{align}
where $D$ means the $x$ component of the $\mathbf{D}$ operator.
The spatial part of $V_{jj'}$ is determined by the usual
dipole moment matrix elements.

Using Eqs.~(\ref{H0elements}) and (\ref{Velements}), we obtain
an eigenvalue equation (\ref{Feigen2}) with a known, infinite matrix
$\tilde{H}.$ For practical calculations \textendash{} like
in all numerical models \textendash{} we have to truncate this matrix,
and use a sufficiently large, but finite part of it, which is determined
by the initial state as well as the strength of the laser-atom interaction
(that is characterized by $E_{0}$). States corresponding to the \char`\"{}static\char`\"{}
eigenfunctions $|n,l,m\rangle=|n,l,m;k=0\rangle$ should necessarily
be taken into account. The number of nonzero $k$ values (that
are directly related to the high harmonics) playing an important role
in the time evolution, depends on $E_{0}.$ (E.g., for a weak excitation
corresponding to linear response, $k=\pm1$ suffices.) From the experimental
point of view, the initial states $|\psi\rangle(0)$ that can be prepared
in the most straightforward manner are the hydrogen-like bound states
$|n,l,m\rangle$ or dipole coupled superpositions of them. Since generally
downward transitions $n\rightarrow n'<n$ cannot be excluded, the
ground state $|n=1,l=0,m=0\rangle$ always has to be taken into account.
As a rule of thumb, the highest the initial value of $n$ is, the more elements of the
numerically necessary finite basis has (and the problem
is numerically more expensive to solve).

\bigskip

On the other hand, we have to recall, that hydrogen-like bound states
$|n,l,m\rangle$, which are orthogonal to the positive-energy states
in the continuum, does not form a complete basis. Additionally, the
usual picture of the process of HHG in gas samples (emission of an
electron, its motion in the laser field and recombination with the
parent ion with the emission of the energy that were gained during
this process as high harmonic radiation) strongly involves the continuum.
Therefore, in order to describe the problem appropriately, we have
to use a different set of spatial functions. The Sturmian states \cite{Avery2006},
the elements of which in coordinate representation can be written
as
\begin{align}
|S_{n,l,m}^{\alpha}\rangle(r,\theta,\phi) & =\frac{\alpha^{3/2}2^{l+1}e^{-\alpha r}}{(2l+1)!} \sqrt{\frac{(l+n)!}{n(-l+n-1)!}} \label{Sturmian wavefunction}\\
 & \times(\alpha r)^{l}{}_{1}\mathrm{F}_{1}(l-n+1;2l+2;2\alpha r)\mathrm{Y}_{l}^{m}(\theta,\phi)\nonumber
\end{align}
form a basis in the space of normalizable states for all values of
$\alpha$ (we use $\alpha=1$), meaning an optimal choice for a norm-preserving,
unitary dynamics that involves the continuum as well. By multiplying
$|S_{n,l,m}^{\alpha}\rangle$ by $e^{ik\omega t},$ we obtain an analogue
of the states $|n,l,m;k\rangle$ that is suitable for the description
of the process of HHG. All the machinery described for $|n,l,m;k\rangle$
can be repeated also for $|S_{n,l,m}^{\alpha}\rangle e^{ik\omega t},$
but we have to keep in mind that the orthogonality relation for the
Sturmian functions involves a weight function of $1/r$ \cite{Avery2006,Popelier2011a}, thus the
inner product (\ref{innerprod}) has to be replaced by
\begin{equation}
\widetilde{\langle\varphi_{1}|\varphi_{2}\rangle} =\int dt\int\frac{1}{r}d^{3}\mathbf{r}\varphi_{1}^{*}(\mathbf{r},t)\varphi_{2}(\mathbf{r},t),\label{innerprodS}
\end{equation}
and the matrix elements of $H_{0}$ and $V(t)$ has to be evaluated
using this inner product.

\section{Results}
\label{resultsec}
%%%%%%%%%%%%%%%%%%%%%%%%%%%%%%%%%%%%%%%%%%%%%%%%%%%%%%%%%%%%%%%%%%%%%%%%%%%%%%%%%%%%%%%%%%%%%%%%%%%%%%%%%%%%%%
\begin{figure}
\begin{center}
\includegraphics[width=8cm]{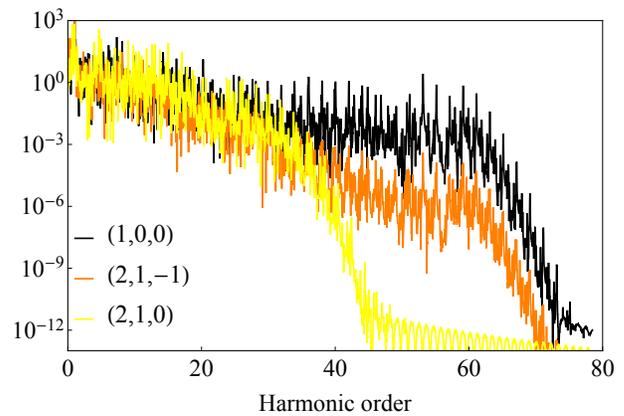}
\caption{HHG spectra for different initial sates $|n,l,m\rangle.$ The quantum numbers are indicated in the legend. The excitation is assumed to be monochromatic, with $E_0=0.1 a.u.$ and $\omega=0.057 a.u.$ (which corresponds to $\lambda=800nm$).}
\label{purefig}
\end{center}
\end{figure}
Although the dynamics of HHG is conveniently described using the Sturmian
states (\ref{Sturmian wavefunction}), the initial states appearing
in Eq.~(\ref{initsate}) that can most easily prepared by conventional,
long, (almost) resonant pulses are energy eigenstates, or dipole coupled
superpositions of them. For the sake of simplicity, in the following
we consider
\begin{equation}
|\psi\rangle(0)=\cos\beta|n=1,l=0,m=0\rangle+e^{i\delta}\sin{\beta}|n'=2,l'=1,m'\rangle,
\label{superpos}
\end{equation}
where $m'=0,\pm1.$ Note that the case of $\beta=0$ ($\pi/2$) means
pure initial energy eigenstates. When $\beta$ equals none of these
values, we have a superposition of the ground state and an
excited state, with a quantum mechanical phase given by $\delta.$
Additionally, when $m'=0,$ selection rules tell us that the external
field that is polarized in the $x$ direction, cannot induce dipole
transitions between the constituents of the initial superpositions,
while for $m'=\pm1,$ this transition is dipole-allowed.
(Note that in order to prepare a superpositions with $m'=0$ in Eq.~(\ref{superpos}), one can use a pulse polarized along the $z$ axis.)
\begin{figure}
\begin{center}
\includegraphics[width=8cm]{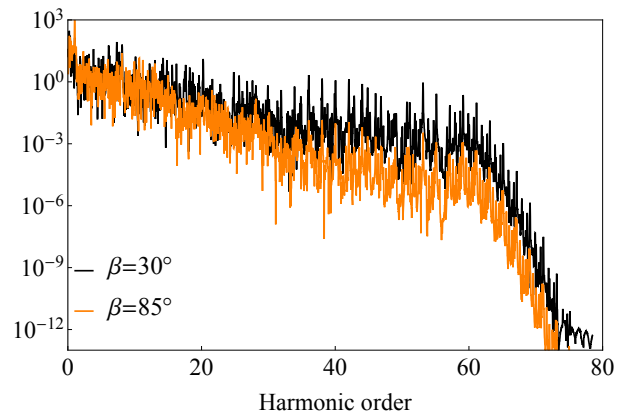}
\caption{HHG spectra for initial superpositions $\cos\beta|n=1,l=0,m=0\rangle+\sin{\beta}|n'=2,l'=1,m'=-1\rangle$ for different values of $\beta.$  The parameters of the exciting field are the same as in Fig.~\ref{purefig}.}
\label{betafig1}
\end{center}
\end{figure}

\bigskip

First, let us consider the case when, instead of a superposition, we have a single energy eigenstate at $t=0.$ Fig.~\ref{purefig} shows the harmonic spectra for the cases when $|\psi\rangle(0)$ is equal to $|n=1,l=0,m=0\rangle,$ $|n=2,l=1,m=0\rangle$ and $|n=2,l=1,m=-1\rangle.$ Let us recall that the cutoff energy can be estimated as $E_c=I_p+3.17$ $U_p,$ where $U_p=e^2E_0^2/4m\omega_0^2$ is the ponderomotive energy. $I_p$ is the ionization potential for the ground state ($|\psi\rangle(0)=|1,0,0\rangle$). As a generalization, one might think that for excited states, $I_p$ should be replaced by $\tilde{I}_p,$ the energy difference between the initial state and the limit of the continuum. This is exactly what we can see in Fig.~\ref{purefig} for $|\psi\rangle(0)=|2,1,0\rangle.$ However, for $|\psi\rangle(0)=|2,1,-1\rangle,$ we do not see a definite decrease of the cutoff frequency. The reason for this is the fact that the state $|2,1,-1\rangle$ is dipole coupled to the ground state, thus during the time evolution the state $|1,0,0\rangle$ also gets populated. In other words, for $|\psi\rangle(0)=|2,1,-1\rangle,$ it is not the initial state the energy difference of which and the continuum determines the value of $\tilde{I}_p.$ In fact, since a "downward" transition to the ground state is dipole-allowed, the value of $\tilde{I}_p$ is undetermined, the dynamics populate various energy eigenstates, among which the ground state has the lowest energy. This is not the case for $|\psi\rangle(0)=|2,1,0\rangle,$ since in this case there is no dipole-allowed transition to the ground state, and $\tilde{I}_p=I_p/2^2 (\approx 13.6 eV/4$ for hydrogen).
\begin{figure}
\begin{center}
\includegraphics[width=8cm]{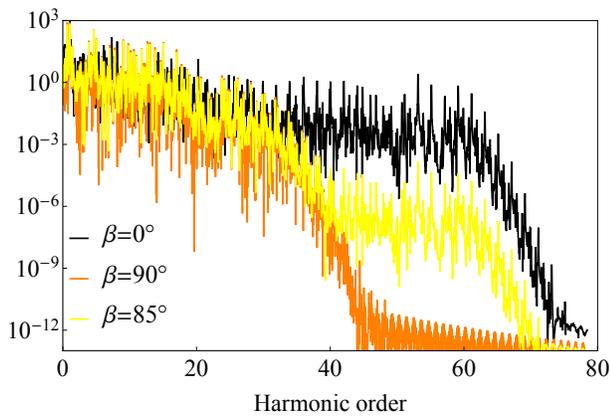}
\caption{HHG spectra for initial superpositions $\cos\beta|n=1,l=0,m=0\rangle+\sin{\beta}|n'=2,l'=1,m'=0\rangle$ for the values of $\beta$ shown by the legend, and $\delta=0.$ The parameters of the exciting field are the same as in Fig.~\ref{purefig}.}
\label{betafig0}
\end{center}
\end{figure}

\bigskip

Having understood the physical picture behind  Fig.~\ref{purefig}, we can conclude that our model can reproduce the most important features of the atomic HHG spectra. Therefore we can turn to the case when the initial states are proper superposition, i.e., $\beta$ is neither $0$ nor $\pi/2$ in Eq.~(\ref{superpos}).
Figs.~\ref{betafig1} and \ref{betafig0} show HHG spectra for different superpositions with $\delta=0$ in Eq.~(\ref{superpos}). For Fig.~\ref{betafig1}, the constituents of the superpositions are dipole coupled, unlike for Figs.~\ref{betafig0}, where $\langle 1,0,0|D|2,1,0\rangle=0.$ The difference between the figures is clear: the effect of the initial atomic coherence is much stronger for the case of Fig.~\ref{betafig0}. As we have discussed above, for $|\psi\rangle(0)=|2,1,-1\rangle,$ the dynamics is unavoidably coupled to the ground state $|1,0,0\rangle.$ Along this line, the combination of these two states as the initial superposition is not expected to strongly depend on the weight of the constituents, since these two states get mixed during the time evolution anyway.

On the other hand, when the constituents of the initial superposition are not dipole coupled, their dynamics follow separated "ladders" to the continuum and back again. This independence leads to the double cutoff structure seen in Fig.~\ref{betafig0}, where both of the cutoffs belonging to $|1,0,0\rangle$ and $|2,1,0\rangle$ are clearly visible when the initial state is the superposition of these states. (Note that  $|2,1,0\rangle$  has to be dominant in order to see this structure, since when the weight of  $|1,0,0\rangle$ is large enough, it produces observable signal also in the region between the "two cutoffs" and the double step-like behaviour of the spectra becomes less apparent on a logarithmic scale that is traditionally used for these plots.) Let us also note that a similar double cutoff structure was obtained in Ref.~\cite{W96} using a different numerical approach. That is, although the constituents of the superposition of the states $|1,0,0\rangle$ and $|2,1,0\rangle$ essentially evolves as two independent states, their weights (determined by $\beta$ in Eq.\ref{superpos}) create observable differences in the spectra.

\bigskip

\begin{figure}
\begin{center}
\includegraphics[width=8cm]{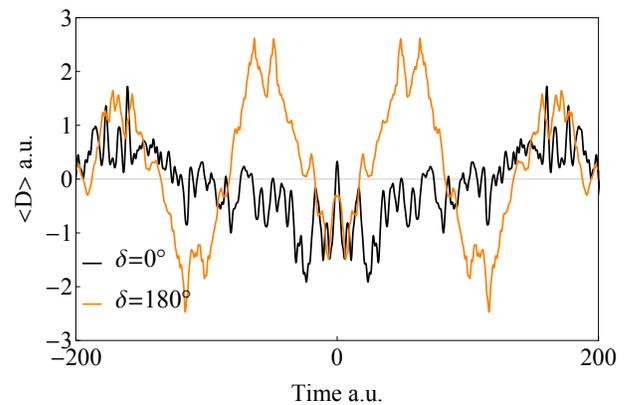}
\caption{The time dependence of the HHG signal for a superposition $\cos\beta|n=1,l=0,m=0\rangle+\sin{\beta}|n'=2,l'=1,m'=-1\rangle$ with $\beta=\pi/4$ and the indicated values of $\delta.$ The parameters of the exciting field are the same as in Fig.~\ref{purefig}.}
\label{timedepfig}
\end{center}
\end{figure}
%%%%%%%%%%%%%%%%%%%%%%%%%%%%%%%%%%%%%%%%%%%%%%%%%%%%%%%%%%%%%%%%%%%%%%%%%%%%%%%%%%%%%%%%%%%%%%%%%%%%%%%%%%%%%%%
The most sophisticated signature of quantum mechanical atomic coherence is the dependence of the dynamics on the relative phase $\delta$ that appears in Eq.~(\ref{superpos}). However, there is no $\delta$ dependence (neither in the spectra, nor in the time evolution of the dipole moment expectation values) for the superpositions of states $|1,0,0\rangle$ and $|2,1,0\rangle.$ In order to understand this, we have to recall the constituents of these superpositions follow independent dynamics, and the dipole moment matrix element between them always vanishes. Therefore both $|1,0,0\rangle$ and $|2,1,0\rangle$ produce their HHG signal independently, and, in this sense, although the term $e^{i\delta}$ in Eq.~(\ref{superpos}) multiplies only $|2,1,0\rangle,$ it plays the role of an overall, irrelevant phase factor. This is exactly what we have seen in our simulations.

On the other hand, for dipole coupled states, there is a nonzero cross term $\langle 1,0,0|D|2,1,-1\rangle$ already at $t=0,$ and its nontrivial time evolution gives an observable contribution to the HHG spectra, as well as to the time dependent HHG signals. As an example, Fig.~\ref{timedepfig} shows the time dependent dipole moment expectation value for the same value of $\beta,$ but for different phases $\delta$ in a superposition of these states.  As we can see, the question whether the sign  between the constituents of the initial  superposition is $+$ or $-$ ($\delta=0$ or $\pi$ in the figure), plays a decisive role in the time evolution of the system. Generally, the positions, the heights as well as the widths of the peaks do depend on the value of $\delta,$ which can be controlled experimentally. The substantially different HHG signals seen in Fig.~\ref{timedepfig} point towards the control of high harmonic radiation by utilizing atomic coherence effects.

\section{Summary}
\label{summarysec}
As it is known, Floquet theorem is an efficient method to solve the time-dependent Schr\"{o}dinger equation for monochromatic
excitation. Having expanded the corresponding equations in a Sturmian basis, we obtained a method that is particularly suitable for the description
of high-order harmonic generation (HHG) in hydrogen-like gases. The quantum mechanical coherence of the initial states were found to have strong influence on the HHG
spectra. In more details, by considering superpositions of usual hydrogenic energy eigenstates as initial states, we have shown that the weights of the constituent states play an important role in the production of the high harmonic radiation. Moreover, for a linearly polarized exciting field, when the constituents of the initial superposition are dipole-coupled, the time-dependent HHG signal is shown to be also sensitive to the quantum mechanical phase difference between the constituents of the initial superpositions. This demonstrates that atomic coherence effects can control the properties of high harmonic radiation.

\bigskip

The ELI-ALPS project (Grants No.  GOP-1.1.1-12/B-2012-000  and  No.  GINOP-2.3.6-15-2015-00001)  is  supported  by  the  European  Union  and  co-financed by the European Regional Development Fund.
Our work was also supported by the European Social Fund under contract EFOP-3.6.2-16-2017-00005.
% Bibliography
%\bibliographystyle{plain}

% Full bibliography added automatically for Optics Letters submissions; the following line will simply be ignored if submitting to other journals.
% Note that this extra page will not count against page length
%\bibliographyfullrefs{sample}

%Manual citation list
%\begin{thebibliography}{1}
%\bibitem{Zhang:14}
%Y.~Zhang, S.~Qiao, L.~Sun, Q.~W. Shi, W.~Huang, %L.~Li, and Z.~Yang,
% \enquote{Photoinduced active terahertz metamaterials with nanostructured
%vanadium dioxide film deposited by sol-gel method,} Opt. Express \textbf{22},
%11070--11078 (2014).
%\end{thebibliography}

\begin{thebibliography}{10}
\newcommand{\enquote}[1]{``#1''}

\bibitem{KI09}
F.~Krausz and M.~Ivanov, \enquote{Attosecond physics,} Rev. Mod. Phys.
  \textbf{81}, 163--234 (2009).

\bibitem{FarkasT92}
G.~Farkas and C.~T\'{o}th, \enquote{Proposal for attosecond light pulse
  generation using laser induced multiple-harmonic conversion processes in rare
  gases,} Physics Letters A \textbf{168}, 447 -- 450 (1992).

\bibitem{Agostini2004}
P.~Agostini and L.~F. DiMauro, \enquote{The physics of attosecond light
  pulses,} Reports on Progress in Physics \textbf{67}, 813 (2004).

\bibitem{McPherson87}
A.~McPherson, G.~Gibson, H.~Jara, U.~Johann, T.~S. Luk, I.~A. McIntyre,
  K.~Boyer, and C.~K. Rhodes, \enquote{Studies of multiphoton production of
  vacuum-ultraviolet radiation in the rare gases,} J. Opt. Soc. Am. B
  \textbf{4}, 595--601 (1987).

\bibitem{F88}
M.~Ferray, A.~L'Huillier, X.~F. Li, L.~A. Lompre, G.~Mainfray, and C.~Manus,
  \enquote{Multiple-harmonic conversion of 1064 nm radiation in rare gases,}
  Journal of Physics B: Atomic, Molecular and Optical Physics \textbf{21}, L31
  (1988).

\bibitem{TG09}
U.~Teubner and P.~Gibbon, \enquote{High-order harmonics from laser-irradiated
  plasma surfaces,} Rev. Mod. Phys. \textbf{81}, 445--479 (2009).

\bibitem{Tsakiris2006}
G.~D. Tsakiris, K.~Eidmann, J.~{Meyer-ter-Vehn}, and F.~Krausz, \enquote{Route
  to intense single attosecond pulses,} New Journal of Physics \textbf{8}, 19
  (2006).

\bibitem{Kahaly2013}
S.~Kahaly, S.~Monchoc\'e, H.~Vincenti, T.~Dzelzainis, B.~Dromey, M.~Zepf,
  P.~Martin, and F.~Qu\'er\'e, \enquote{Direct observation of density-gradient
  effects in harmonic generation from plasma mirrors,} Phys. Rev. Lett.
  \textbf{110}, 175001 (2013).

\bibitem{Baeva2006}
T.~Baeva, S.~Gordienko, and A.~Pukhov, \enquote{Theory of high-order harmonic
  generation in relativistic laser interaction with overdense plasma,} Phys.
  Rev. E \textbf{74}, 046404 (2006).

\bibitem{Ghimire10}
S.~Ghimire, A.~D. DiChiara, E.~Sistrunk, P.~Agostini, L.~F. DiMauro, and D.~A.
  Reis, \enquote{Observation of high-order harmonic generation in a bulk
  crystal,} Nature Physics \textbf{7}, 138--141 (2011).

\bibitem{Ciappina2014}
M.~F. Ciappina, J.~A. P\'erez-Hern\'andez, T.~Shaaran, M.~Lewenstein,
  M.~Kr\"uger, and P.~Hommelhoff, \enquote{High-order-harmonic generation
  driven by metal nanotip photoemission: Theory and simulations,} Phys. Rev. A
  \textbf{89}, 013409 (2014).

\bibitem{0953-4075-24-15-004}
A.~L'Huillier, K.~J. Schafer, and K.~C. Kulander, \enquote{Theoretical aspects
  of intense field harmonic generation,} Journal of Physics B: Atomic,
  Molecular and Optical Physics \textbf{24}, 3315 (1991).

\bibitem{C93b}
P.~B. Corkum, \enquote{Plasma perspective on strong field multiphoton
  ionization,} Phys. Rev. Lett. \textbf{71}, 1994--1997 (1993).

\bibitem{Krause1992}
J.~L. Krause, K.~J. Schafer, and K.~C. Kulander, \enquote{Calculation of
  photoemission from atoms subject to intense laser fields,} Phys. Rev. A
  \textbf{45}, 4998--5010 (1992).

\bibitem{levenstein}
M.~Lewenstein, P.~Balcou, M.~Y. Ivanov, A.~L'Huillier, and P.~B. Corkum,
  \enquote{Theory of high-harmonic generation by low-frequency laser fields,}
  Phys. Rev. A \textbf{49}, 2117--2132 (1994).

\bibitem{F883}
G.~Floquet, \enquote{Sur les équations différentielles linéaires a
  coefficients périodiques,} Ann. \'{E}cole Norm. Sup. \textbf{12}, 46--88
  (1883).

\bibitem{PS89}
R.~M. Potvliege and R.~Shakeshaft, \enquote{Multiphoton processes in an intense
  laser field: Harmonic generation and total ionization rates for atomic
  hydrogen,} Phys. Rev. A \textbf{40}, 3061--3079 (1989).

\bibitem{FK97}
F.~H.~M. Faisal and J.~Z. Kami\ifmmode~\acute{n}\else \'{n}\fi{}ski,
  \enquote{Floquet-bloch theory of high-harmonic generation in periodic
  structures,} Phys. Rev. A \textbf{56}, 748--762 (1997).

\bibitem{SPF15}
V.~Szaszk\'o-Bog\'ar, F.~M. Peeters, and P.~F\"oldi, \enquote{Oscillating
  spin-orbit interaction in two-dimensional superlattices: Sharp transmission
  resonances and time-dependent spin-polarized currents,} Phys. Rev. B
  \textbf{91}, 235311 (2015).

\bibitem{Goscinski}
O.~Goscinski, \enquote{Preliminary research report no. 217, quantum chemistry
  group, uppsala university, (1968),} .

\bibitem{Goscinski2002}
O.~{Goscinski}, Advances in Quantum Chemistry \textbf{41}, 57--85 (2002).

\bibitem{joachain2012atoms}
C.~Joachain, N.~Kylstra, and R.~Potvliege, \emph{Atoms in Intense Laser Fields}
  (Cambridge University Press, 2012).

\bibitem{ayadi}
V.~Ayadi, M.~G.Benedict, P.~Dombi, and P.~F\"{o}ldi, \enquote{Atomic coherence
  effects in few-cycle pulse induced ionization,} Eur. Phys. J. D (2016) 70:
  266.  (2016).

\bibitem{Avery2006}
J.~Avery and J.~Avery, \emph{Generalized Sturmians and Atomic Spectra} (World
  Scientific, 2006).

\bibitem{Popelier2011a}
P.~Popelier, \emph{Solving the Schr\"{o}dinger Equation: Has Everything Been
  Tried?} (Imperial College Press, 2011).

\bibitem{W96}
J.~B. Watson, A.~Sanpera, X.~Chen, and K.~Burnett, \enquote{Harmonic generation
  from a coherent superposition of states,} Phys. Rev. A \textbf{53},
  R1962--R1965 (1996).

\end{thebibliography}
\end{document}